\documentstyle[floats,aps]{revtex}
\begin{document}
\renewcommand{\thefootnote}{\fnsymbol{footnote}}
\draft
\title{\large\bf 
  An open-boundary integrable model of three coupled $XY$ spin chains}

\author{Anthony J. Bracken,  Xiang-Yu Ge \footnote{E-mail:
	    xg@maths.uq.edu.au}, 
        Yao-Zhong Zhang  \footnote {Queen Elizabeth II Fellow.
                                   E-mail: yzz@maths.uq.edu.au}
             and 
        Huan-Qiang Zhou \footnote {On leave of absence from Dept of
	         Physics, Chongqing University, Chongqing 630044, China.
                 E-mail: hqzhou@cqu.edu.cn}} 

\address{Department of Mathematics,University of Queensland,
		     Brisbane, Qld 4072, Australia}

\maketitle

\vspace{10pt}

\begin{abstract}
The integrable open-boundary conditions for the model of three coupled
one-dimensional $XY$ spin chains are considered in the framework of
the quantum inverse scattering method. 
The diagonal boundary K-matrices are found and 
a class of integrable boundary terms is determined.
The boundary model Hamiltonian is solved by using
the coordinate space Bethe ansatz technique and Bethe ansatz equations
are derived.
\end{abstract}




\def\a{\alpha}
\def\b{\beta}
\def\d{\delta}
\def\e{\epsilon}
\def\g{\gamma}
\def\k{\kappa}
\def\l{u}
\def\m{v}
\def\o{\omega}
\def\t{\theta}
\def\s{\sigma}
\def\D{\Delta}
\def\L{\Lambda}


\def\beq{\begin{equation}}
\def\eeq{\end{equation}}
\def\bea{\begin{eqnarray}}
\def\eea{\end{eqnarray}}
\def\ba{\begin{array}}
\def\ea{\end{array}}
\def\no{\nonumber}
\def\le{\langle}
\def\re{\rangle}
\def\lt{\left}
\def\rt{\right}

\newcommand{\sect}[1]{\setcounter{equation}{0}\section{#1}}
\renewcommand{\theequation}{\thesection.\arabic{equation}}
\newcommand{\reff}[1]{eq.~(\ref{#1})}

\vskip.3in

\sect{Introduction\label{int}}

Many systems in nature have boundaries. Therefore the
problem of how to extend a bulk integrable model to 
incorporate integrable open-boundary conditions
becomes very important. In general, boundary conditions spoil
the integrability of the bulk system and thus boundary models are
not generally integrable. 

One of the recent developments in the theory of
completely integrable lattice systems with boundaries
is Sklyanin's work \cite{Skl88} on the boundary quantum inverse
scattering method (QISM). This work is extended
by Mezinzescu et al \cite{Mez91} to treat boundary systems with
the R-matrices satisfying the less restrictive condition of
crossing-symmetry. It is noted by de Vega et al \cite{deV93} that
if R-matrices satisfy a weaker version of
crossing-symmetry, called crossing-unitarity,
the corresponding boundary systems can still be
treated by the approach of Mezinzescu et al.
Boundary integrable lattice systems 
have subsequently been extensively investigated in the literature
(see, e.g. \cite{Yun95,Zho96a,Asa96,Zho96b,Zha97,Bra97}).

In many interesting cases
the R-matrices enjoy neither $PT$ invariance and crossing-symmetry
nor any kind of weaker versions such as crossing-unitarity. Two
non-trivial examples are R-matrices \cite{Zho96R}
corresponding to the models of two
coupled and three coupled one-dimensional $XY$ spin chains introduced in
\cite{Bar91}. For such systems,
the formalism of \cite{Skl88,Mez91}
is not applicable.
In \cite{Zho96b} (\cite{Bra97}), a very general (graded) reflection
equation (RE) algebra 
has been proposed and the corresponding (supersymmetric) boundary
QISM has been formulated. The formalism applies to any (supersymmetric)
lattice boundary system where an invertible R-matrix exists.

The aim of this paper is to study integrable open-boundary
conditions for the model of three coupled one dimensional $XY$ spin chains.
The quantum integrability of the bulk model has been established in
\cite{Zho96R} by deriving the explicit form of the quantum R-matrix.
It is seen that this R-matrix
does not possess crossing-unitarity.
So we apply the generalized scheme developed in \cite{Zho96b,Bra97}, which
is reformulated in section II.
We are interested in constructing 
diagonal boundary K-matrices. Since the R-matrix for the model
concerned is a
$64\times 64$ matrix with 216 non-zero elements, the whole procedure
of solving REs for K-matrices
is very involved. Furthermore, due to the absence of crossing-unitarity,
there is no isomorphism between the two K-matrices $K_\pm$ and the
two REs have to be solved separately. This fact is in contrast to
the known cases in the literature, where $K_+$ can be derived from
$K_-$ by means of the isomorphism betweem them. 
In the present case,
however, deriving $K_+$ is highly non-trivial, as it also involves
determining the inverse of the partially transposed R-matrix.
We list our results for the K-matrices in section III and more details
can be found in the Appendices A and B. The explicit formulae for $K_\pm(u)$
verify that the two K-matrices are not related by automorphisms.
In section IV we use these K-matrices 
to determine the open-boundary model Hamiltonian, which turns out to be
a complicated algebraic manipulation. The reason is that at the zero
spectral-parameter point, the traces of the K-matrix
and of its 1st and 2nd derivatives are all equal to zero, and  thus
we have to expand the boundary transfer matrix
to the 4-th order in the spectral parameter.
The Hamiltonian of the boundary model is extracted from
the 4-th derivative of the boundary transfer matrix. 
We then solve the boundary model by means of the coordinate
Bethe ansatz method and derive the Bethe ansatz equations.
In Appendix A, we sketch the procedure
of solving the RE for $K_-(u)$,  and  in Appendix B we list the
216 non-zero matrix elements of $\tilde{R}(u)$ appearing in the RE
for $K_+(u)$. The construction of $K_+(u)$ is then similar to that of 
$K_-(u)$. As a sideline, we show in Appendix C 
that all R-matrices associated with finite
dimensional representations of a quantum affine algebra enjoy the
crossing-unitarity property. We emphasize, however, that the
R-matrix used in the present paper does not belong to this class.

\sect{Reflection Equations and the Boundary Transfer Matrix \label{for}}

Let $V$ be a finite-dimensional linear space. Let the operator-valued
function $R:C\rightarrow End(V\otimes V)$ be a solution to the quantum
Yang-Baxter equation (QYBE)
\beq
R_{12}(u_1,u_2)R_{13}(u_1,u_3)R_{23}(u_2,u_3)=R_{23}(u_2,u_3)
R_{13}(u_1,u_3)R_{12}(u_1,u_2).
\eeq
Here $R_{jk}(u)$ denotes the matrix on $V\otimes V\otimes V$ acting on the
$j$-th and $k$-th spaces and as an identity on the remaining space. The
variables $u_1,\;u_2$ and $u_3$ are  spectral parameters.
No assumption that the spectral parameters are additive is to be made. Let $P$
be the permutation operator in $V\otimes V$, i.e.,
$P(x\otimes y)=y\otimes x,~\forall x,y\in V$.
Then $R_{21}(u)=P_{12}R_{12}(u)P_{12}$. 
With the help of the R-matrix $R_{jk}(u)$, we form the monodromy matrix
$T(u)$ for an $L$-site spin chain, 
\beq
T(u)=L_{0L}(u)\cdots L_{01}(u)
\eeq
where the subscript $0$ labels
the auxiliary space $V$
and  $L_{0j}(u) \equiv R_{0j}(u,0)$. Indeed, one can 
show that $T(u)$ generates  a representation of the quantum Yang-Baxter
algebra,
\beq
R_{12}(u_1,u_2)\stackrel {1}{T}(u_1) \stackrel {2}{T}(u_2)=\stackrel {2}{T}(u_2)
\stackrel {1}{T}(u_1) R_{12}(u_1,u_2).\label{yb-alg}
\eeq
where $\stackrel {1}{X} \equiv  X \otimes 1$ and
$\stackrel {2}{X} \equiv  1 \otimes X$,
for any matrix $ X \in End(V) $.

In order to construct open-boundary integrable spin chains, we 
introduce the following REs which the so-called boundary K-matrices
satisfy:
\bea
R_{12}(u_1,u_2)\stackrel {1} {K}_-(u_1) R_{21}(u_2,-u_1)
\stackrel {2}{K}_-(u_2)&=& 
\stackrel {2}{K}_-(u_2) R_{12}(u_1,-u_2)
\stackrel {1}{K}_-(u_1) R_{21}(-u_2,-u_1),\no\\ 
R_{21}^{t_1t_2}(u_2,u_1)\stackrel {1}{K^{t_1}_+}(u_1) {\tilde R}_{12}(-u_1,u_2)
\stackrel {2}{K^{t_2}_+}(u_2)& = &
\stackrel {2}{K^{t_2}_+}(u_2) {\tilde R}_{21}(-u_2,u_1)
\stackrel {1}{K^{t_1}_+}(u_1) R_{12}^{t_1t_2}(-u_1,-u_2),\label{REs}    
\eea
where we have defined a new object ${\tilde R}$ through the relation
\beq
{\tilde R} _{21}^{t_1} (-u_2,u_1) R^{t_2}_{12}(u_1,-u_2) =1\label{cross}
\eeq
and $t_i$ stands for the transposition taken in the $i$-th space.
In all cases,
quantum R-matrices enjoy the unitarity property \footnote{One can always
normalize the R-matrices so that the right hand side of
(\ref{unitarity}) is equal to the identity, as shown.},
\beq
 R_{12}(u_1,u_2)R_{21}(u_2,u_1) = 1.\label{unitarity}
\eeq

We now show that the second RE in
(\ref{REs}) is indeed the correct ``conjugation" to the first one, so
that the boundary transfer matrices defined as usual constitute a commuting
family. Following  Sklyanin's arguments \cite{Skl88}, one
may show that the quantity ${\cal T}_-(u)$ given by
\beq
{\cal T}_-(u) = T(u) {K}_-(u) T^{-1}(-u) 
\eeq
satisfies the same relation as $K_-(u)$:
\beq
R_{12}(u_1,u_2)\stackrel {1} {\cal T}_-(u_1) R_{21}(u_2,-u_1)
\stackrel {2}{\cal T}_-(u_2)= 
\stackrel {2}{\cal T}_-(u_2) R_{12}(u_1,-u_2)
\stackrel {1}{\cal T}_-(u_1) R_{21}(-u_2,-u_1).\label{re-for-t}
\eeq
Thus if one defines the boundary transfer matrix as
\beq
t(u) = tr (K_+(u){\cal T}_-(u))=tr\lt(K_+(u)T(u)K_-(u)T^{-1}(-u)\rt),
\eeq
where $tr$ denotes the trace taken over the auxiliary space $V$,
then it can be shown \cite{Bra97} that
\beq
[t(u_1),t(u_2)] = 0.
\eeq

The REs (\ref{REs}) are generalizations of those introduced 
by Sklyanin \cite{Skl88} and Mezincescu et al \cite{Mez91}. In contrast
to those works, we do not impose any constraint conditions on the
R-matrix. Therefore the REs (\ref{REs})
apply to any bosonic (or non-supersymmetric) case where an invertible R-matrix
exists.

\sect{Boundary K-matrices for the coupled spin chain model\label{Boun}}

We consider a spin chain model defined by the following Hamiltonian
\beq
H = \sum ^{L-1}_{j=1} H _{j,j+1}+B_L +B_R,\label{hamiltonian}
\eeq
where $H_{j,j+1}$ denotes the bulk Hamiltonian density of
three $XY$ spin chains coupled to each other \cite{Bar91}
\bea
H_{j,j+1} = \sum _{\alpha}\;(\sigma^+_{j(\alpha)}\sigma^-_{j+1(\alpha)}+
\sigma^-_{j(\alpha)}\sigma^+_{j+1(\alpha)})
\exp[\eta \sum _{\alpha' \neq \alpha }
\sigma ^+_{j+\t(\alpha'-\alpha)(\alpha')}
\sigma ^-_{j+\t(\alpha'-\alpha)(\alpha')}].\label{bulk-h}
\eea
Here $\sigma ^{\pm}_{j(\alpha)}= \frac {1}{2}(\sigma ^x_{j(\alpha)}
\pm i\sigma ^y_{j(\alpha)}) $, with $\sigma ^x_{j(\alpha)},
\sigma ^y_{j(\alpha)},\sigma ^z_{j(\alpha)}$ being the usual Pauli 
spin operators at site $j$  corresponding
to the $\a$-th ($\a=1,2,3$) $XY$ spin chain;
$\t(\a'-\a)$ is a step
function of $(\a'-\a)$; and $\eta$ is a coupling constant.
The left and right boundary terms $B_L$ and $B_R$ have the form
\bea
B_L&=&\frac {1}{2c_- \exp(2\eta)}[\cosh ^2 \eta 
	  \sum _{\a}\sigma ^z_{1(\a)} 
          +\frac {\sinh \eta \cosh \eta}{2} \sum _{\a \neq \b}
	  \sigma ^z_{1(\a)} \s ^z_{1(\b)}+\sinh^2 \eta\; \s ^z_{1(1)}
	  \s ^z_{1(2)} \s^z _{1(3)}], \no\\ 
B_R&=&\frac {c_+ \exp(6\eta)}{2}[\cosh ^2 \eta 
	  \sum _{\a}\sigma ^z_{L(\a)} 
          +\frac {\sinh \eta \cosh \eta}{2} \sum _{\a \neq \b}
	  \sigma ^z_{L(\a)} \s ^z_{L(\b)}+\sinh^2 \eta\; \s ^z_{L(1)}
	  \s ^z_{L(2)} \s ^z_{L(3)}],  
\eea
where $c_\pm$ are parameters describing the boundary effects.
After a generalized Jordan-Wigner transformation, the
Hamiltonian (\ref{hamiltonian}) becomes
a strongly correlated electronic system with boundary interactions.
If one restricts the Hilbert space to the
one which only consists of, say, $\s _{(1)},~\s _{(2)}$, then 
Hamiltonian ({\ref{hamiltonian}) reduces to that of two coupled
$XY$ open chains with special boundary interactions, which has been
considered in \cite{Zho96b}.

We shall establish the quantum integrability of the system defined by
the Hamiltonian (\ref{hamiltonian}), by using the general formalism 
described in the previous section. Let us first of all
recall some basic results for the bulk model (\ref{bulk-h})
with the periodic boundary conditions. As was shown in
\cite{Zho96R},
the bulk model Hamiltonian  commutes with  a one-parameter 
family of bulk transfer matrices $\tau(u)$
of a two-dimensional lattice statistical mechanics model.
This transfer matrix is the trace of a monodromy
matrix $T(u)$ with $L_{0j}(u)$ of the form,
\beq
L_{0j}(u) = L^{(1)}_{0j}(u) L^{(2)}_{0j}(u) L^{(3)}_{0j}(u),\label{mono}
\eeq
where
\bea
L^{(\alpha)}_{0j}(u) &=& \frac{1}{2}(1+\sigma ^z_{j(\alpha)}
\sigma ^z_{0(\alpha)}) + \frac{1}{2} u (1-\sigma ^z_{j(\alpha)}
\sigma ^z_{0(\alpha)})
\exp(\eta \sum^{3}_{\stackrel {\alpha'=1}{\alpha' \neq \alpha}}
\sigma ^+_{0(\alpha')}
\sigma ^-_{0(\alpha')})\no\\
& &+(\sigma^-_{j(\alpha)} \sigma^+_{0(\alpha)} + \sigma ^+_{j(\alpha)}
\sigma ^-_{0(\alpha)}) \sqrt {1+
\exp(2 \eta \sum^{3}_{\stackrel {\alpha'=1}{\alpha' \neq \alpha}}
\sigma ^+_{0(\alpha')}\sigma ^-_{0(\alpha')}) u ^2}. 
\eea
The commutativity of the bulk transfer matrices
$\tau(u)$ for different values of the spectral parameter $ u$ follows
from the fact that
the $L$ matrix $L_{0j}(u)$ (\ref{mono}) satisfies the Yang-Baxter 
algebra (\ref{yb-alg}). The explicit form of the corresponding R-matrix 
$R_{12}(u_1,u_2)$  can be found in the third reference in \cite{Zho96R}.
Here we only emphasize that the local monodromy matrix as well as 
the quantum R-matrix
does not possess crossing symmetry and crossing-unitarity. 

In order to describe integrable systems with boundary conditions different
from the periodic ones, let us first of all solve the REs for the two boundary
K-matrices $K_{\pm}(u)$. As mentioned in the Introduction,
there is no isomorphism between $K_-(u)$ and
$K_+(u)$, and thus we have to solve the two REs separately.
For our purpose, we only look for solutions where $K_\pm(u)$ are
diagonal. After complicated algebraic manipulations (for
a sketch of the whole procedure, see Appendix A), we find
\beq
K_-(u) = \frac {1}{\exp (6\eta) c_-^3}
\left ( \begin {array} {cccccccc}
A_-(u)  &0&0&0&0&0&0&0\\
0&B_-(u) &0&0&0&0&0&0 \\
0&0&B_-(u) &0&0&0&0&0\\
0&0&0&C_-(u)&0&0&0&0\\
0&0&0&0&B_-(u)&0&0&0\\
0&0&0&0&0&C_-(u)&0&0\\
0&0&0&0&0&0&C_-(u)&0\\
0&0&0&0&0&0&0&D_-(u)
\end {array}  \right ),
\eeq
with
\bea
A_-(u)&=&(c_-+u)(e^{2\eta} c_- +u)(e^{4\eta} c_-+u),\no\\
B_-(u)&=&(c_--u)(e^{2\eta} c_- +u)(e^{4\eta}c_-+u),\no\\
C_-(u)&=&(c_--u)(e^{2\eta} c_--u)(e^{4\eta} c_-+u),\no\\
D_-(u)&=&(c_--u)(e^{2\eta} c_- -u)(e^{4\eta} c_--u).\no
\eea
It is much more tedious to find the boundary K-matrix $K_+(u)$, since the
corresponding RE is more involved. We list the final result here and
some details can be found in Appendices B and A,
\beq
K_+(u)= \left ( \begin {array} {cccccccc}
A_+(u)  &0&0&0&0&0&0&0\\
0&B_+(u) &0&0&0&0&0&0 \\
0&0&C_+(u) &0&0&0&0&0\\
0&0&0&D_+(u)&0&0&0&0\\
0&0&0&0&E_+(u)&0&0&0\\
0&0&0&0&0&F_+(u)&0&0\\
0&0&0&0&0&0&G_+(u)&0\\
0&0&0&0&0&0&0&H_+(u)
\end {array}  \right ) ,  \label{k+} 
\eeq
with
\bea
A_+(u)&=&e^{6\eta}(c_++u)(e^{2\eta} c_+ +u)
         (e^{4\eta}c_++u) ,\no\\
B_+(u)&=&e^{6\eta}(-e^{4\eta} c_++u)(e^{2\eta} c_+ +u)
         (e^{4\eta}c_++u) ,\no\\
C_+(u)&=&e^{4\eta}(c_++u)(-e^{4\eta} c_++u)(e^{4\eta} c_+ +u),\no\\
D_+(u)&=&e^{4\eta}(-e^{4\eta} c_++u)(-e^{6\eta} c_+ +u)
         (e^{4\eta} c_++u) ,\no\\
E_+(u)&=&e^{2\eta}(e^{2\eta}c_++u)(-e^{4\eta} c_+ +u)
         (e^{4\eta}c_++u) ,\no\\
F_+(u)&=&e^{2\eta}(-e^{4\eta} c_++u)(-e^{6\eta} c_+ +u)
         (e^{4\eta} c_++u) ,\no\\
G_+(u)&=&(-e^{4\eta} c_++u)(-e^{6\eta} c_+ +u)
         (e^{4\eta}c_++u) ,\no\\
H_+(u)&=&(-e^{4\eta} c_++u)(-e^{6\eta} c_+ +u)
         (-e^{8\eta}c_++u).\no 
\eea
 
The above explicit formulae for $K_\pm(u)$, derived by solving the two 
REs separately,
show that no automorphism between $K_+(u)$ and $K_-(u)$ exists 
and $K_+(u)$ cannot be obtained from $K_-(u)$, as claimed.

\sect{Embedding of the Hamiltonian into the boundary transfer matrix
      and the Bethe ansatz equations}

To  show that the Hamiltonian (\ref{hamiltonian}) can be embedded 
into the boundary transfer matrix
$t(u)$ constructed in section II,  is
an involved algebraic manipulation. This is because the traces of $K_+(0)$,
and of its first derivative $K'_+(0)$ and second derivative 
$K''_+(0)$, 
are all equal to zero. So at best we can only  expect 
that the Hamiltonian (\ref{hamiltonian})
appears as the fourth derivative of the boundary transfer matrix with respect
to the spectral parameter $u$, at $u=0$. 

Let us expand the
local monodromy matrix $L_{0j}(u)$ to the fourth order in the
spectral parameter $u$,
\beq
L_{0j}(u)= (1+H_{j0}u +\frac {1}{2!}B_{j0}u^2+\frac {1}{3!} C_{j0}u^3 
+\frac {1}{4!}
D_{j0}u^4+\cdots)L_{0j}(0),
\eeq
where,
\bea
H_{j0}&=&H_{j0}^{(1)}+P_{j0}^{(1)}H_{j0}^{(2)}P_{j0}^{(1)}
+P_{j0}^{(1)}P_{j0}^{(2)}H_{j0}^{(3)}P_{j0}^{(2)}P_{j0}^{(1)},\no\\
B_{j0}&=&B_{j0}^{(1)}+P_{j0}^{(1)}B_{j0}^{(2)}P_{j0}^{(1)}
+P_{j0}^{(1)}P_{j0}^{(2)}B_{j0}^{(3)}P_{j0}^{(2)}P_{j0}^{(1)}
+2H_{j0}^{(1)}P_{j0}^{(1)}H_{j0}^{(2)}P_{j0}^{(1)}\no\\
& &+2H_{j0}^{(1)}P_{j0}^{(1)}P_{j0}^{(2)}H_{j0}^{(3)}P_{j0}^{(2)}P_{j0}^{(1)}
+2P_{j0}^{(1)}H_{j0}^{(2)}P_{j0}^{(1)}
P_{j0}^{(1)}P_{j0}^{(2)}H_{j0}^{(3)}P_{j0}^{(2)}P_{j0}^{(1)},\no\\
C_{j0}&=&3B_{j0}^{(1)}P_{j0}^{(1)}H_{j0}^{(2)}P_{j0}^{(1)}
+3B_{j0}^{(1)}P_{j0}^{(1)}P_{j0}^{(2)}H_{j0}^{(3)}P_{j0}^{(2)}P_{j0}^{(1)}
 \no\\
& &+3H_{j0}^{(1)}P_{j0}^{(1)}B_{j0}^{(2)}P_{j0}^{(1)}
+3H_{j0}^{(1)}P_{j0}^{(1)}P_{j0}^{(2)}B_{j0}^{(3)}P_{j0}^{(2)}P_{j0}^{(1)
  }\no\\ 
& &+3P_{j0}^{(1)}B_{j0}^{(2)}P_{j0}^{(1)}
P_{j0}^{(1)}P_{j0}^{(2)}H_{j0}^{(3)}P_{j0}^{(2)}P_{j0}^{(1)}
+3P_{j0}^{(1)}H_{j0}^{(2)}P_{j0}^{(1)}
P_{j0}^{(1)}P_{j0}^{(2)}B_{j0}^{(3)}P_{j0}^{(2)}P_{j0}^{(1)}\no\\
& &+6H_{j0}^{(1)}P_{j0}^{(1)}H_{j0}^{(2)}P_{j0}^{(1)}
P_{j0}^{(1)}P_{j0}^{(2)}H_{j0}^{(3)}P_{j0}^{(2)}P_{j0}^{(1)}
\eea
with
\bea
H_{j0}^{(\a)}&=&(\s_{j(\a)}^-\s_{0(\a)}^+ +\s_{j(\a)}^+\s_{0(\a)}^-)
  \exp(\eta\sum_{\stackrel{\a'=1}{\a\neq \a'}}^3\;\s_{0(\a')}^+
  \s_{0(\a')}^-),\no\\
P_{j0}^{(\a)}&=&\frac{1+\s_{j(\a)}^z\s_{0(\a)}^z}{2}+
(\s_{j(\a)}^-\s_{0(\a)}^+ +\s_{j(\a)}^+\s_{0(\a)}^-),\no\\
B_{j0}^{(\a)}&=&(\s_{j(\a)}^-\s_{0(\a)}^+ +\s_{j(\a)}^+\s_{0(\a)}^-)
  P_{j0}^{(\a)}
  \exp(2\eta\sum_{\stackrel{\a'=1}{\a\neq \a'}}^3\;\s_{0(\a')}^+
  \s_{0(\a')}^-).
\eea
Substituting the expansion for $L_{0j}(u)$
into the boundary transfer matrix $t(u)$, and after a
lengthy but straightforward algebraic calculation, one finds
\beq
t(u) =  C_1 u^3 + C_2 (H + const.) u^4 + \cdot\cdot\cdot,
\eeq
where $C_i (i = 1,2,\cdot\cdot\cdot)$ are some scalar functions of the boundary
constant $c_+$. It can be shown that the  Hamiltonian (\ref{hamiltonian}) 
is related to the fourth derivative of the boundary transfer matrix
$t(u)$, 
\beq
H \equiv \frac {t^{(4)}(0)}{8\; tr K_+^{'''}(0)}=
  \sum ^{L-1}_{j=1} H_{j,j+1} +\frac {1}{2} \stackrel {1}{K'}_-(0)
  +\frac{3}{trK_+'''(0)}\;tr(K''_+(0)H^2_{L0}),\label{4-th der}
\eeq
where
\beq
H_{j,j+1}=L_{0,j+1}(0)L'_{0j}(0)L^{-1}_{0j}(0)L^{-1}_{0,j+1}(0).
\eeq
In deriving (\ref{4-th der}) we have used the following relations which
hold in the present case,
\bea
&&tr K_+^{(n)}(0) H_{L0} =0,~~~~~~n=0,1,2,3,\no\\
&&tr K_+^{(n)}(0) B_{L0} =0,~~~~~~n=0,1,\no\\
&&tr K_+(0) H_{L0}B_{L0} =0,~~~~~~
  tr K_+(0) B_{L0}H_{L0} =0,~~~~~~
  tr\lt(K_+(0)C_{L0}H_{L0}\rt)=0,\no\\
&&tr K_+(0)H_{L0}^2=0,~~~~~~
  tr K'_+(0)H_{L0}^2=0,~~~~~~
  tr K'_+(0)C_{L0}=0,\no\\
&&tr K_+(0)H^3_{L0}=0,~~~~~~
  tr K'_+(0)H^3_{L0}=0,\no\\
&&tr\lt(K_+(0)H_{L0}C_{L0}\rt)=0,~~~~~
  tr\lt(K_+(0)H_{L0}B_{L0}H_{L0}\rt)=0,\no\\
&&tr\lt(K_+(0)H_{L0}^2B_{L0}\rt)=0,~~~~~~
  tr\lt(K_+(0)H_{L0}^4\rt)=0.
\eea

We have shown that the  Hamiltonian (\ref{hamiltonian}) of the 
boundary model of three coupled one dimensional $XY$ spin
chains is related to a class of commuting transfer matrices.
As a result, the system has an infinite number of higher conserved currents
which are involutive with each other, and therefore
the system under study is completely integrable.

Having established the quantum integrability of the model,
let us  now  solve it by using
the coordinate space Bethe ansatz method. The procedure is similar to
that for other models \cite{Asa96,Zha97,Bra97}.
The Bethe ansatz equations are 
\bea
e^{ik_j2(L+1)}\zeta(k_j;p_1)\zeta(k_j;p_L)
&=&\prod ^{M_1}_{\a =1}\frac {\sin [\frac {1}{2}
(k_j-\Lambda ^{(1)}_{\a})+\frac {i\eta}{2}]}
{\sin [\frac {1}{2}
(k_j-\Lambda ^{(1)}_{\a}) -\frac {i\eta}{2}]}
\frac {\sin [\frac {1}{2}
(k_j+\Lambda ^{(1)}_{\a})+\frac {i\eta}{2}]}
{\sin [\frac {1}{2}
(k_j+\Lambda ^{(1)}_{\a}) -\frac {i\eta}{2}]},\no\\
\prod ^N_{\a =1}\frac {\sin [\frac {1}{2}(\Lambda ^{(1)}_{\g}
  -k_\a)+\frac{i\eta}{2}]}
{\sin [\frac {1}{2}(\Lambda ^{(1)}_{\g}
   -k_\a)-\frac{i\eta}{2}]}
\frac {\sin [\frac {1}{2}(\Lambda ^{(1)}_{\g}
   +k_\a+\frac{i\eta}{2}]}
{\sin [\frac {1}{2}(\Lambda ^{(1)}_{\g}
   +k_\a)-\frac{i\eta}{2}]}
&=&
  \prod ^{M_1}_{\stackrel {\g ' =1}{\g ' \neq \g}}\frac
 {\sin [\frac {1}{2}(\Lambda ^{(1)}_{\g}-\Lambda ^{(1)}_{\g '})
 +i\eta]}{\sin [\frac {1}{2}(\Lambda ^{(1)}_{\g}-\Lambda ^{(1)}_{\g '})
 -i\eta]}\frac {\sin [\frac {1}{2}(\Lambda ^{(1)}_{\g}+\Lambda ^{(1)}_{\g '})
 +i\eta]}{\sin [\frac {1}{2}(\Lambda ^{(1)}_{\g}+\Lambda ^{(1)}_{\g '})
 -i\eta]}\no\\
& &\times\prod ^{M_2}_{\delta=1}
 \frac {\sin [\frac {1}{2}(\Lambda ^{(1)}_{\g}-\lambda ^{(2)}_{\delta})
 -\frac{i\eta}{2}]}
 {\sin [\frac {1}{2}(\Lambda ^{(1)}_{\g}-\lambda ^{(2)}_{\delta})
 +\frac{i\eta}{2}]}
 \frac {\sin [\frac {1}{2}(\Lambda ^{(1)}_{\g}+\lambda ^{(2)}_{\delta})
 -\frac{i\eta}{2}]}
 {\sin [\frac {1}{2}(\Lambda ^{(1)}_{\g}+\lambda ^{(2)}_{\delta})
 +\frac{i\eta}{2}]},\no\\
\prod ^{M_2}_{\stackrel {\g '=1}{\g ' \neq \g}}
\frac {\sin [\frac {1}{2}(\Lambda ^{(2)}_{\g}-\Lambda ^{(2)}_{\g '})
 +i\eta]}{\sin [\frac {1}{2}(\Lambda ^{(2)}_{\g}-\Lambda ^{(2)}_{\g '})
 -i\eta]}\frac {\sin [\frac {1}{2}(\Lambda ^{(2)}_{\g}+\Lambda ^{(2)}_{\g '})
 +i\eta]}{\sin [\frac {1}{2}(\Lambda ^{(2)}_{\g}+\Lambda ^{(2)}_{\g '})
  -i\eta]}
&=&
 \prod ^{M_1}_{\a =1}
 \frac {\sin [\frac {1}{2}(\Lambda ^{(2)}_{\g}-\Lambda ^{(1)}_{\a})
 +\frac{i\eta}{2}]}
 {\sin [\frac {1}{2}(\Lambda ^{(2)}_{\g}-\Lambda ^{(1)}_{\a})
 -\frac{i\eta}{2}]}
 \frac {\sin [\frac {1}{2}(\Lambda ^{(2)}_{\g}+\Lambda ^{(1)}_{\a})
 +\frac{i\eta}{2}]}
 {\sin [\frac {1}{2}(\Lambda ^{(2)}_{\g}+\Lambda ^{(1)}_{\a})
 -\frac{i\eta}{2}]},
\eea
where 
\beq
\zeta (k;p)= (1-pe^{-ik})/(1-pe^{ik}),~~~~
p_1=\frac {1}{c_- \exp (4 \eta)},~~~~
p_L=c_+ \exp (4 \eta).
\eeq
The energy eigenvalue $E$ of the  model is given by
$E=2\sum ^N_{j=1}\cos k_j$ (modulo an unimportant additive constant,
which we drop).

\sect{Conclusion \label{con}}

In this paper, we have studied integrable open-boundary conditions for the
coupled spin chain model. The  quantum integrability
of the boundary system has been established by the fact
that the corresponding Hamiltonian  may be embedded into
a one-parameter family of commuting transfer matrices. Moreover, the Bethe
Ansatz equations are derived by means of the coordinate space Bethe ansatz
approach. This provides us with a basis for computing the finite size
corrections to the low-lying energies in the system, which in turn will
allow
us to use the boundary conformal field theory technique to study
the critical properties of the boundary.

Let us add a few comments. Firstly, it should be emphasized
that the formalism exposed here applies to all bosonic lattice systems where
an invertible R-matrix exists, since one does not impose any constraint
conditions on the quantum R-matrix. Secondly, the
results can be generalized to the case of $s$ coupled $XY$ chains, with
$s$ arbitrary.
It is interesting
to note that in the case $s=2$, the Hamiltonian appears as the third
derivative of the transfer matrix with respect to the spectral parameter
$u$ \cite{Zho96b}, whereas in the present case, $s=3$,
it appears as the fouth derivative.
Therefore, one may
expect that for arbitrary $s$, the corresponding Hamiltonian will
appear as the $(s+1)$-th derivative of the transfer matrix.
Finally, it seems interesting to derive the Bethe ansatz
equations by using the algebraic Bethe ansatz approach.

\vskip.3in
Y.-Z.Z and H.-Q.Z are supported by Australian Research Council, University of
Queensland New Staff Research Grant and External Support Enabling Grant. H.-Q.Z
would like to thank the Department of Mathematics, University of
Queensland for kind hospitality.
He is also partially supported by the National Natural Science Foundation
of China and Sichuan Young Investigators Science and Technology Fund.

\appendix

\sect{Derivation of the boundary K-matrices}

In this appendix, we sketch the procedure of solving the RE for 
$K_-(u)$.
The R-matrix appeared in this paper is an $64\times 64$ matrix with 
216 non-zero elements (see the third reference in
\cite{Zho96R}). We only
consider  diagonal solution. We parametrize
$K_-(u)$ as
\beq
K_-(u) =\left (
\begin {array} {cccccccc}
a(u)  &0&0&0&0&0&0&0\\
0&b(u) &0&0&0&0&0&0 \\
0&0&c(u) &0&0&0&0&0\\
0&0&0&d(u)&0&0&0&0\\
0&0&0&0&e(u)&0&0&0\\
0&0&0&0&0&f(u)&0&0\\
0&0&0&0&0&0&g(u)&0\\
0&0&0&0&0&0&0&h(u)
\end {array}  \right ).
\eeq
The RE for $K_-(u)$ consists of 216 functional 
equations, many of which are
dependent. To solve them, we group these 216 functional equations into four
categories: Category 0 contains 8 identities, each  made
of two terms; Category 1 contains 48 equations, each of 4
terms; Category 2 contains 96 equations, each of 8 terms;
Category 3 contains 64 equations, each of 16 terms. 
Now divide the 48 functional
equations in Category 1 into 12 sets of equations, so that each set
contains 4 equations. It turns out that in any set of the 4 equations,
two of them are satisfied identically and the other two are equivalent
to each other. Therefore in these 48 equations in Category 1, only
12 need to be solved. Picking the following 7 equations
out of those 12 equations,
\bea
&&a(u_1)a(u_2)R_{21,12}(u_1,u_2)R_{12,12}(u_2,-u_1)+
b(u_1)a(u_2)R_{21,21}(u_1,u_2)R_{21,12}(u_2,-u_1)\no\\
&&~= a(u_1)b(u_2)R_{21,12}(u_1,-u_2)R_{12,12}(-u_2,-u_1)+
b(u_1)b(u_2)R_{21,21}(u_1,-u_2)R_{21,12}(-u_2,-u_1),\\
&&a(u_1)a(u_2)R_{31,13}(u_1,u_2)R_{13,13}(u_2,-u_1)+
c(u_1)a(u_2)R_{31,31}(u_1,u_2)R_{31,13}(u_2,-u_1)\no\\
&&~= a(u_1)c(u_2)R_{31,13}(u_1,-u_2)R_{13,13}(-u_2,-u_1)+
c(u_1)c(u_2)R_{31,31}(u_1,-u_2)R_{31,13}(-u_2,-u_1),\\
&&a(u_1)a(u_2)R_{51,15}(u_1,u_2)R_{15,15}(u_2,-u_1)+
e(u_1)a(u_2)R_{51,51}(u_1,u_2)R_{51,15}(u_2,-u_1)\no\\
&&~= a(u_1)e(u_2)R_{51,15}(u_1,-u_2)R_{15,15}(-u_2,-u_1)+
e(u_1)e(u_2)R_{51,51}(u_1,-u_2)R_{51,15}(-u_2,-u_1),\\
&&b (u_1)b(u_2)R_{42,24}(u_1,u_2)R_{24,24}(u_2,-u_1)+
d(u_1)b(u_2)R_{42,42}(u_1,u_2)R_{42,24}(u_2,-u_1)\no\\
&&~= b(u_1)d(u_2)R_{42,24}(u_1,-u_2)R_{24,24}(-u_2,-u_1)+
d(u_1)d(u_2)R_{42,42}(u_1,-u_2)R_{42,24}(-u_2,-u_1),\\
&&b(u_1)b(u_2)R_{62,26}(u_1,u_2)R_{26,26}(u_2,-u_1)+
f(u_1)b(u_2)R_{62,62}(u_1,u_2)R_{62,26}(u_2,-u_1)\no\\
&&~= b(u_1)f(u_2)R_{62,26}(u_1,-u_2)R_{26,26}(-u_2,-u_1)+
f(u_1)f(u_2)R_{62,62}(u_1,-u_2)R_{62,26}(-u_2,-u_1),\\
&&c(u_1)c(u_2)R_{73,37}(u_1,u_2)R_{37,37}(u_2,-u_1)+
g(u_1)c(u_2)R_{73,73}(u_1,u_2)R_{73,37}(u_2,-u_1)\no\\
&&~= c(u_1)g(u_2)R_{73,37}(u_1,-u_2)R_{37,37}(-u_2,-u_1)+
g(u_1)g(u_2)R_{73,73}(u_1,-u_2)R_{73,37}(-u_2,-u_1),\\
&&d(u_1)d(u_2)R_{84,48}(u_1,u_2)R_{48,48}(u_2,-u_1)+
h(u_1)d(u_2)R_{84,84}(u_1,u_2)R_{84,48}(u_2,-u_1)\no\\
&&~= d(u_1)h(u_2)R_{84,48}(u_1,-u_2)R_{48,48}(-u_2,-u_1)+
h(u_1)h(u_2)R_{84,84}(u_1,-u_2)R_{84,48}(-u_2,-u_1),
\eea
and using some nontrivial tricks of variable separation, we determine
7 constants, $c_1,\;c_2,\;\cdots,\;c_7$, by the relations
\bea
&&\frac{b(u)}{a(u)}=\frac{c_1-u}{c_1+u},~~~~~
  \frac{c(u)}{a(u)}=\frac{c_2-u}{c_2+u},~~~~~
  \frac{e(u)}{a(u)}=\frac{c_3-u}{c_3+u},\no\\
&&\frac{d(u)}{a(u)}=\frac{c_1-u}{c_1+u}\cdot\frac{c_4-u}{c_4+u},~~~~~
  \frac{f(u)}{a(u)}=\frac{c_1-u}{c_1+u}\cdot\frac{c_5-u}{c_5+u},\no\\
&&\frac{g(u)}{a(u)}=\frac{c_2-u}{c_2+u}\cdot\frac{c_6-u}{c_6+u},~~~~~
  \frac{h(u)}{a(u)}=\frac{c_1-u}{c_1+u}\cdot\frac{c_4-u}{c_4+u}
  \cdot\frac{c_7-u}{c_7+u}.\label{7-c}
\eea

Then divide the 96 equations in Category 2 into 6 sets of equations.
Each set contains 16 equations. In any set of the 16 equations, eight 
equations are identities and only two of the remaining 8 equations
are relevant. Therefore out of these 96 equations in
Category 2, only 12 equations can provide useful information. 
Inserting (\ref{7-c}) into the following 4 equations picked from those 12
equations,
\bea
&&a(u_1)a(u_2)R_{41,14}(u_1,u_2)R_{14,14}(u_2,-u_1)+
b(u_1)a(u_2)R_{41,23}(u_1,u_2)R_{23,14}(u_2,-u_1)\no\\
&&~~~~+ c(u_1)a(u_2)R_{41,32}(u_1,u_2)R_{32,14}(u_2,-u_1)+
d(u_1)a(u_2)R_{41,41}(u_1,u_2)R_{41,14}(u_2,-u_1)\no\\
&&~=a(u_1)d(u_2)R_{41,14}(u_1,-u_2)R_{14,14}(-u_2,-u_1)+
b(u_1)d(u_2)R_{41,23}(u_1,-u_2)R_{23,14}(-u_2,-u_1)\no\\
&&~~~~+ c(u_1)d(u_2)R_{41,32}(u_1,-u_2)R_{32,14}(-u_2,-u_1)+
d(u_1)d(u_2)R_{41,41}(u_1,-u_2)R_{41,14}(-u_2,-u_1),\\
&&a(u_1)a(u_2)R_{61,16}(u_1,u_2)R_{16,16}(u_2,-u_1)+
b(u_1)a(u_2)R_{61,25}(u_1,u_2)R_{25,16}(u_2,-u_1)\no\\
&&~~~~+ e(u_1)a(u_2)R_{61,52}(u_1,u_2)R_{52,16}(u_2,-u_1)+
f(u_1)a(u_2)R_{61,61}(u_1,u_2)R_{61,16}(u_2,-u_1)\no\\
&&~=a(u_1)f(u_2)R_{61,16}(u_1,-u_2)R_{16,16}(-u_2,-u_1)+
b(u_1)f(u_2)R_{61,25}(u_1,-u_2)R_{25,16}(-u_2,-u_1)\no\\
&&~~~~+ e(u_1)f(u_2)R_{61,52}(u_1,-u_2)R_{52,16}(-u_2,-u_1)+
f(u_1)f(u_2)R_{61,61}(u_1,-u_2)R_{61,16}(-u_2,-u_1),\\
&&a(u_1)a(u_2)R_{71,17}(u_1,u_2)R_{17,17}(u_2,-u_1)+
c(u_1)a(u_2)R_{71,35}(u_1,u_2)R_{35,17}(u_2,-u_1)\no\\
&&~~~~+ e(u_1)a(u_2)R_{71,53}(u_1,u_2)R_{53,17}(u_2,-u_1)+
g(u_1)a(u_2)R_{71,71}(u_1,u_2)R_{71,17}(u_2,-u_1)\no\\
&&~=a(u_1)g(u_2)R_{71,17}(u_1,-u_2)R_{17,17}(-u_2,-u_1)+
c(u_1)g(u_2)R_{71,35}(u_1,-u_2)R_{35,17}(-u_2,-u_1)\no\\
&&~~~~+ e(u_1)g(u_2)R_{71,53}(u_1,-u_2)R_{53,17}(-u_2,-u_1)+
g(u_1)g(u_2)R_{71,71}(u_1,-u_2)R_{71,17}(-u_2,-u_1),\\
&&b(u_1)b(u_2)R_{82,28}(u_1,u_2)R_{28,28}(u_2,-u_1)+
d(u_1)b(u_2)R_{82,46}(u_1,u_2)R_{46,28}(u_2,-u_1)\no\\
&&~~~~+ f(u_1)b(u_2)R_{82,64}(u_1,u_2)R_{64,28}(u_2,-u_1)+
h(u_1)b(u_2)R_{82,82}(u_1,u_2)R_{82,28}(u_2,-u_1)\no\\
&&~=b(u_1)h(u_2)R_{82,28}(u_1,-u_2)R_{28,28}(-u_2,-u_1)+
d(u_1)h(u_2)R_{82,46}(u_1,-u_2)R_{46,28}(-u_2,-u_1)\no\\
&&~~~~+ f(u_1)h(u_2)R_{82,64}(u_1,-u_2)R_{64,28}(-u_2,-u_1)+
h(u_1)h(u_2)R_{82,82}(u_1,-u_2)R_{82,28}(-u_2,-u_1),
\eea
one can determine the following relationships among the 7 constants:
\beq
c_2=c_3=c_1,~~~~c_4=c_5=c_6=e^{2\eta}c_1,~~~~c_7=e^{4\eta}c_1.
\eeq
So we have uniquely determined the ratios $\frac{b(u)}{a(u)}$ {\it etc.} up to
a free parameter $c_1\equiv c_-$. Finally it can be checked that the
64 functional equations in Category 3 are identically satisfied with the
$\frac{b(u)}{a(u)}$ {\it etc.}
determined as above. So we end up with the diagonal
solution $K_-(u)$ in section III
with one arbitrary parameter satisfying all 216 functional
equations.

\sect{The matrix elements of ${\tilde R}$}

The first step towards solving the RE for $K_+(u)$ is to find
$\tilde{R}(\l,\m)$ defined by (\ref{cross}). After long algebraic
computations, one finds that the non-zero matrix
elements of $\tilde{R}(\l,\m)$ are
\bea
{\tilde R}_{11,11}(\l ,\m )&=&
-\frac {(1+\l \m)(1+e^{2\eta}\l \m)
(1+e^{4\eta}\l \m)^2
(1+e^{6\eta}\l \m)
(1+e^{8\eta}\l \m)}
{e^{6\eta}(\l- \m)^2(e^{2\eta}\l- \m)
(\l-e^{2\eta}\m)
(e^{4\eta}\l -\m)
(\l-e^{4\eta}\m)},\no\\
{\tilde R}_{12,12}(\l,\m)&=&
{\tilde R}_{13,13}(\l,\m)=
{\tilde R}_{15,15}(\l,\m)=
\frac {(1+\l \m)(1+e^{2\eta}\l \m)
(1+e^{4\eta}\l \m)^2
(1+e^{6\eta}\l \m)}
{e^{4\eta}(\l- \m)^2(e^{2\eta}\l- \m)
(\l-e^{2\eta}\m)
(\l-e^{4\eta}\m)},\no\\
{\tilde R}_{12,21}(\l,\m)&=&
{\tilde R}_{51,15}(\l,\m)=
e^{2\eta}{\tilde R}_{13,31}(\l,\m)=
e^{2\eta}{\tilde R}_{31,13}(\l,\m)=
e^{4\eta}{\tilde R}_{15,51}(\l,\m)=
e^{4\eta}{\tilde R}_{21,12}(\l,\m)\no\\
&=&\frac {(1+\l \m)(1+e^{2\eta}\l \m)
(1+e^{4\eta}\l \m)^2
(1+e^{6\eta}\l \m)
\sqrt{(1+e^{4\eta}\l ^2)(1+e^{4\eta}\m ^2)}}
{e^{2\eta}(\l- \m)^2(e^{2\eta}\l- \m)
(\l-e^{2\eta}\m)
(e^{4\eta}\l -\m)
(\l-e^{4\eta}\m)},\no\\
{\tilde R}_{14,14}(\l,\m)&=&
{\tilde R}_{16,16}(\l,\m)=
{\tilde R}_{17,17}(\l,\m)=
-\frac {(1+\l \m)(1+e^{2\eta}\l \m)
(1+e^{4\eta}\l \m)^2}
{e^{2\eta}(\l- \m)^2
(\l-e^{2\eta}\m)
(\l-e^{4\eta}\m)},\no\\
{\tilde R}_{14,23}(\l,\m)&=&
{\tilde R}_{16,25}(\l,\m)=
{\tilde R}_{52,16}(\l,\m)=
{\tilde R}_{53,17}(\l,\m)\no\\
&=&
e^{\eta}{\tilde R}_{14,32}(\l,\m)=
e^{\eta}{\tilde R}_{35,17}(\l,\m)
=e^{2\eta}{\tilde R}_{17,35}(\l,\m)=
e^{2\eta}{\tilde R}_{32,14}(\l,\m)\no\\
&=&
e^{3\eta}{\tilde R}_{16,52}(\l,\m)=
e^{3\eta}{\tilde R}_{17,53}(\l,\m)=
e^{3\eta}{\tilde R}_{23,14}(\l,\m)
=e^{3\eta}{\tilde R}_{25,16}(\l,\m)\no\\
&=&-\frac {(1+\l \m)(1+e^{2\eta}\l \m)
(1+e^{4\eta}\l \m)^2
\sqrt{(1+e^{2\eta}\l ^2)(1+e^{4\eta}\m ^2)}}
{e^{\eta}(\l- \m)^2(e^{2\eta}\l- \m)
(\l-e^{2\eta}\m)
(\l-e^{4\eta}\m)},\no\\
{\tilde R}_{14,41}(\l,\m)&=&
{\tilde R}_{71,17}(\l,\m)=
e^{2\eta}{\tilde R}_{16,61}(\l,\m)=
e^{2\eta}{\tilde R}_{61,16}(\l,\m)=
e^{4\eta}{\tilde R}_{17,71}(\l,\m)=
e^{4\eta}{\tilde R}_{41,14}(\l,\m)\no\\
&=&-\frac {(1+\l \m)(1+e^{2\eta}\l \m)
(1+e^{4\eta}\l \m)^2
\sqrt{(1+e^{2\eta}\l ^2)(1+e^{4\eta}\l ^2)
(1+e^{2\eta}\m ^2)(1+e^{4\eta}\m ^2)}}
{(\l- \m)^2(e^{2\eta}\l- \m)
(\l-e^{2\eta}\m)
(e^{4\eta}\l -\m)
(\l-e^{4\eta}\m)},\no\\
{\tilde R}_{18,18}(\l,\m)&=&
\frac {(1+\l \m)(1+e^{2\eta}\l \m)
(1+e^{4\eta}\l \m)}
{(\l- \m)
(\l-e^{2\eta}\m)
(\l-e^{4\eta}\m)},\no\\
{\tilde R}_{18,27}(\l,\m)&=&
{\tilde R}_{54,18}(\l,\m)=
e^{\eta}{\tilde R}_{18,36}(\l,\m)=
e^{\eta}{\tilde R}_{36,18}(\l,\m)=
e^{2\eta}{\tilde R}_{18,54}(\l,\m)=
e^{2\eta}{\tilde R}_{27,18}(\l,\m)\no\\
&=&\frac {(1+\l \m)(1+e^{2\eta}\l \m)
(1+e^{4\eta}\l \m)
\sqrt{(1+\l ^2)(1+e^{4\eta}\m ^2)}}
{(\l- \m)^2
(\l-e^{2\eta}\m)
(\l-e^{4\eta}\m)},\no\\
{\tilde R}_{18,45}(\l,\m)&=&
{\tilde R}_{72,18}(\l,\m)=
e^{\eta}{\tilde R}_{18,63}(\l,\m)=
e^{\eta}{\tilde R}_{63,18}(\l,\m)=
e^{2\eta}{\tilde R}_{18,72}(\l,\m)=
e^{2\eta}{\tilde R}_{45,18}(\l,\m)\no\\
&=&\frac {(1+\l \m)(1+e^{2\eta}\l \m)
(1+e^{4\eta}\l \m)
\sqrt{(1+\l ^2)(1+e^{2\eta}\l ^2)(1+e^{2\eta}\m ^2)(1+e^{4\eta}\m ^2)}}
{(\l- \m)^2(e^{2\eta}\l- \m)
(\l-e^{2\eta}\m)
(\l-e^{4\eta}\m)},\no\\
{\tilde R}_{18,81}(\l,\m)&=&
{\tilde R}_{81,18}(\l,\m)
=\frac {(1+\l \m)(1+e^{2\eta}\l \m)
(1+e^{4\eta}\l \m)}
{(\l- \m)^2(e^{2\eta}\l- \m)
(\l-e^{2\eta}\m)
(e^{4\eta}\l -\m)
(\l-e^{4\eta}\m)}\no\\
& &\times \sqrt{(1+\l ^2)(1+e^{2\eta}\l ^2)(1+e^{4\eta}
 \l ^2)(1+\m ^2)(1+e^{2\eta}\m ^2)
 (1+e^{4\eta}\m ^2)},\no\\
{\tilde R}_{21,21}(\l,\m)&=&
{\tilde R}_{31,31}(\l,\m)=
{\tilde R}_{51,51}(\l,\m)=
\frac {(1+\l \m)(1+e^{2\eta}\l \m)
(1+e^{4\eta}\l \m)^2
(1+e^{6\eta}\l \m)}
{e^{4\eta}(\l- \m)^2(e^{2\eta}\l- \m)
(\l-e^{2\eta}\m)
(e^{4\eta}\l -\m)},\no\\
{\tilde R}_{22,22}(\l,\m)&=&
{\tilde R}_{33,33}(\l,\m)=
{\tilde R}_{55,55}(\l,\m)=
-\frac {(1+\l \m)^2(1+e^{2\eta}\l \m)
(1+e^{4\eta}\l \m)^2
(1+e^{6\eta}\l \m)}
{e^{2\eta}(\l- \m)^2(e^{2\eta}\l- \m)
(\l-e^{2\eta}\m)
(e^{4\eta}\l -\m)
(\l-e^{4\eta}\m)},\no\\
{\tilde R}_{23,23}(\l,\m)&=&
{\tilde R}_{25,25}(\l,\m)=
{\tilde R}_{32,32}(\l,\m)=
{\tilde R}_{35,35}(\l,\m)=
{\tilde R}_{52,52}(\l,\m)=
{\tilde R}_{53,53}(\l,\m)\no\\
&=&-\frac {(1+\l \m)(1+e^{2\eta}\l \m)
(1+e^{4\eta}\l \m)^2}
{e^{3\eta}(\l- \m)^2(e^{2\eta}\l- \m)
(\l-e^{2\eta}\m)},\no\\
{\tilde R}_{23,32}(\l,\m)&=&
{\tilde R}_{35,53}(\l,\m)=
e^{\eta}{\tilde R}_{25,52}(\l,\m)\no\\
&=&-\frac {(1+\l \m)(1+e^{2\eta}\l \m)
(1+e^{4\eta}\l \m)^2
(e^{2\eta}(1+e^{2\eta}\l ^2)
(1+e^{2\eta}\m ^2)+(1-e^{2\eta})(1+e^{6\eta}\l \m))}
{e^{4\eta}(\l- \m)^2(e^{2\eta}\l- \m)
(\l-e^{2\eta}\m)
(e^{4\eta}\l -\m)
(\l-e^{4\eta}\m)},\no\\
{\tilde R}_{23,41}(\l,\m)&=&
{\tilde R}_{71,53}(\l,\m)=
e^{-\eta}{\tilde R}_{32,41}(\l,\m)=
e^{-\eta}{\tilde R}_{52,61}(\l,\m)\no\\
&= &e^{-\eta}{\tilde R}_{61,25}(\l,\m)=
e^{-\eta}{\tilde R}_{71,35}(\l,\m)=
e^{\eta}{\tilde R}_{41,23}(\l,\m)=
e^{\eta}{\tilde R}_{53,71}(\l,\m)\no\\
&=&e^{2\eta}{\tilde R}_{25,61}(\l,\m)=
e^{2\eta}{\tilde R}_{35,71}(\l,\m)=
e^{2\eta}{\tilde R}_{41,32}(\l,\m)=
e^{2\eta}{\tilde R}_{61,52}(\l,\m)\no\\
&=&-\frac {(1+\l \m)(1+e^{2\eta}\l \m)
(1+e^{4\eta}\l \m)^2
\sqrt{(1+e^{4\eta}\l ^2)(1+e^{2\eta}\m ^2)}}
{e^{2\eta}(\l- \m)^2(e^{2\eta}\l- \m)
(\l-e^{2\eta}\m)
(e^{4\eta}\l -\m)
},\no\\
{\tilde R}_{24,24}(\l,\m)&=&
{\tilde R}_{26,26}(\l,\m)=
{\tilde R}_{34,34}(\l,\m)=
{\tilde R}_{37,37}(\l,\m)=
{\tilde R}_{56,56}(\l,\m)=
{\tilde R}_{57,57}(\l,\m)\no\\
&=&\frac {(1+\l \m)^2(1+e^{2\eta}\l \m)
(1+e^{4\eta}\l \m)^2}
{e^{\eta}(\l- \m)^2(e^{2\eta}\l- \m)
(\l-e^{2\eta}\m)
(\l-e^{4\eta}\m )
},\no\\
{\tilde R}_{24,42}(\l,\m)&=&
{\tilde R}_{42,24}(\l,\m)=
{\tilde R}_{57,75}(\l,\m)=
{\tilde R}_{75,57}(\l,\m)\no\\
&=&e^{2\eta}{\tilde R}_{26,62}(\l,\m)=
e^{2\eta}{\tilde R}_{37,73}(\l,\m)=
e^{2\eta}{\tilde R}_{43,34}(\l,\m)=
e^{2\eta}{\tilde R}_{65,56}(\l,\m)\no\\
&=&e^{-2\eta}{\tilde R}_{34,43}(\l,\m)=
e^{-2\eta}{\tilde R}_{56,65}(\l,\m)=
e^{-2\eta}{\tilde R}_{62,26}(\l,\m)=
e^{-2\eta}{\tilde R}_{73,37}(\l,\m)\no\\
&=&\frac {(1+\l \m)^2(1+e^{2\eta}\l \m)
(1+e^{4\eta}\l \m)^2
\sqrt{(1+e^{2\eta}\l ^2)(1+e^{2\eta}\m ^2)}}
{(\l- \m)^2(e^{2\eta}\l- \m)
(\l-e^{2\eta}\m)
(e^{4\eta}\l -\m)
(\l-e^{4\eta}\m)},\no\\
{\tilde R}_{27,27}(\l,\m)&=&
{\tilde R}_{36,36}(\l,\m)=
{\tilde R}_{54,54}(\l,\m)=
\frac {(1+\l \m)(1+e^{2\eta}\l \m)
(1+e^{4\eta}\l \m)}
{e^{2\eta}(\l- \m)^2
(\l-e^{2\eta}\m)
},\no\\
{\tilde R}_{27,36}(\l,\m)&=&
{\tilde R}_{36,54}(\l,\m)=
e^{\eta}{\tilde R}_{27,54}(\l,\m)\no\\
&=&\frac {(1+\l \m)(1+e^{2\eta}\l \m)
(1+e^{4\eta}\l \m)
((1+\l \m)
(1+e^{4\eta}\l \m )+(e^{2\eta}\l- \m)(\l- e^{4\eta} \m))}
{e^{3\eta}(\l- \m)^2(e^{2\eta}\l- \m)
(\l-e^{2\eta}\m)
(\l-e^{4\eta}\m)},\no\\
{\tilde R}_{27,45}(\l,\m)&=&
{\tilde R}_{45,27}(\l,\m)=
{\tilde R}_{54,72}(\l,\m)=
{\tilde R}_{72,54}(\l,\m)\no\\
&=&e^{\eta}{\tilde R}_{27,63}(\l,\m)=
e^{\eta}{\tilde R}_{36,72}(\l,\m)=
e^{\eta}{\tilde R}_{45,36}(\l,\m)=
e^{\eta}{\tilde R}_{63,54}(\l,\m)\no\\
&=&e^{-\eta}{\tilde R}_{36,45}(\l,\m)=
e^{-\eta}{\tilde R}_{54,63}(\l,\m)=
e^{-\eta}{\tilde R}_{63,27}(\l,\m)=
e^{-\eta}{\tilde R}_{72,36}(\l,\m)\no\\
&=&\frac {(1+\l \m)(1+e^{2\eta}\l \m)
(1+e^{4\eta}\l \m)
\sqrt{(1+e^{2\eta}\l ^2)(1+e^{2\eta}\m ^2)}}
{e^{2\eta}(\l- \m)^2(e^{2\eta}\l- \m)
(\l-e^{2\eta}\m)
},\no\\
{\tilde R}_{27,72}(\l,\m)&=&
{\tilde R}_{45,54}(\l,\m)=
\sqrt{(1+e^{2\eta}\l ^2)(1+e^{2\eta}\m ^2)}\no\\
& &\times\frac {(1+\l \m)(1+e^{2\eta}\l \m)
(1+e^{4\eta}\l \m)
((1+e^{4\eta}\l ^2)
(1+e^{4\eta}\m ^2)+e^{4\eta}\l \m(1-e^{4\eta})(1+\l \m))}
{e^{4\eta}(\l- \m)^2(e^{2\eta}\l- \m)
(\l-e^{2\eta}\m)
(e^{4\eta}\l -\m)
(\l-e^{4\eta}\m)},\no\\
{\tilde R}_{27,81}(\l,\m)&=&
{\tilde R}_{81,54}(\l,\m)=
e^{-\eta}{\tilde R}_{36,81}(\l,\m)=
e^{-\eta}{\tilde R}_{81,36}(\l,\m)=
e^{-2\eta}{\tilde R}_{54,81}(\l,\m)=
e^{-2\eta}{\tilde R}_{81,27}(\l,\m)\no\\
&=&\frac {(1+\l \m)(1+e^{2\eta}\l \m)
(1+e^{4\eta}\l \m)
\sqrt{(1+e^{2\eta}\l ^2)(1+e^{4\eta}\l ^2)(1+\m ^2)(1+e^{2\eta}\m ^2)}}
{e^{2\eta}(\l- \m)^2(e^{2\eta}\l- \m)
(\l-e^{2\eta}\m)
(e^{4\eta}\l -\m)
},\no\\
{\tilde R}_{28,28}(\l,\m)&=&
{\tilde R}_{38,38}(\l,\m)=
{\tilde R}_{58,58}(\l,\m)=
-\frac {(1+\l \m)^2(1+e^{2\eta}\l \m)
(1+e^{4\eta}\l \m)}
{(\l- \m)^2
(\l-e^{2\eta}\m)
(\l-e^{4\eta}\m)
},\no\\
{\tilde R}_{28,46}(\l,\m)&=&
{\tilde R}_{76,58}(\l,\m)=
e^{\eta}{\tilde R}_{28,64}(\l,\m)=
e^{\eta}{\tilde R}_{38,74}(\l,\m)\no\\
&=&e^{\eta}{\tilde R}_{47,38}(\l,\m)=
e^{\eta}{\tilde R}_{67,58}(\l,\m)=
e^{-\eta}{\tilde R}_{46,28}(\l,\m)=
e^{-\eta}{\tilde R}_{58,76}(\l,\m)\no\\
&=&e^{-2\eta}{\tilde R}_{38,47}(\l,\m)=
e^{-2\eta}{\tilde R}_{58,67}(\l,\m)=
e^{-2\eta}{\tilde R}_{64,28}(\l,\m)=
e^{-2\eta}{\tilde R}_{74,38}(\l,\m)\no\\
&=&-\frac {(1+\l \m)^2(1+e^{2\eta}\l \m)
(1+e^{4\eta}\l \m)
\sqrt{(1+\l ^2)(1+e^{2\eta}\m ^2)}}
{(\l- \m)^2
(\l-e^{2\eta}\m)
(e^{2\eta}\l-\m)
(\l-e^{4\eta}\m)},\no\\
{\tilde R}_{28,82}(\l,\m)&=&
{\tilde R}_{85,58}(\l,\m)=
e^{-2\eta}{\tilde R}_{38,83}(\l,\m)=
e^{-2\eta}{\tilde R}_{83,38}(\l,\m)
=e^{-4\eta}{\tilde R}_{58,85}(\l,\m)=
e^{-4\eta}{\tilde R}_{82,28}(\l,\m)\no\\
&=&-\frac {(1+\l \m)^2(1+e^{2\eta}\l \m)
(1+e^{4\eta}\l \m)
\sqrt{(1+\l ^2)(1+e^{2\eta}\l ^2)(1+\m ^2)(1+e^{2\eta}\m ^2)}}
{(\l- \m)^2(e^{2\eta}\l- \m)
(\l-e^{2\eta}\m)
(e^{4\eta}\l -\m)
(\l-e^{4\eta}\m)},\no\\
{\tilde R}_{32,23}(\l,\m)&=&
{\tilde R}_{53,35}(\l,\m)=
e^{-2\eta}{\tilde R}_{52,25}(\l,\m)\no\\
&=&-\frac {(1+\l \m)(1+e^{2\eta}\l \m)
(1+e^{4\eta}\l \m)^2
(e^{2\eta}(1+e^{2\eta}\l ^2)
(1+e^{2\eta}\m ^2)+\l \m(e^{2\eta}-1)(1+e^{6\eta}\l \m))}
{e^{2\eta}(\l- \m)^2(e^{2\eta}\l- \m)
(\l-e^{2\eta}\m)
(e^{4\eta}\l -\m)
(\l-e^{4\eta}\m)},\no\\
{\tilde R}_{36,27}(\l,\m)&=&
{\tilde R}_{54,36}(\l,\m)=
e^{-\eta}{\tilde R}_{54,27}(\l,\m)\no\\
&=&\frac {(1+\l \m)(1+e^{2\eta}\l \m)
(1+e^{4\eta}\l \m)
(e^{2\eta}(1+\l \m)
(1+e^{4\eta}\l \m )+(e^{2\eta}\l- \m)(\l- e^{4\eta} \m))}
{e^{\eta}(\l- \m)^2(e^{2\eta}\l- \m)
(\l-e^{2\eta}\m)
(\l-e^{4\eta}\m)},\no\\
{\tilde R}_{36,63}(\l,\m)&=&
{\tilde R}_{63,36}(\l,\m)=
\sqrt{(1+e^{2\eta}\l ^2)(1+e^{2\eta}\m ^2)}\no\\
& &\times\frac {(1+\l \m)(1+e^{2\eta}\l \m)
(1+e^{4\eta}\l \m)
[e^{2\eta}(1+e^{2\eta}\l ^2)
(1+e^{2\eta}\m ^2)-\l \m(1-e^{2\eta})(1-e^{6\eta})]}
{e^{2\eta}(\l- \m)^2(e^{2\eta}\l- \m)
(\l-e^{2\eta}\m)
(e^{4\eta}\l -\m)
(\l-e^{4\eta}\m)},\no\\
{\tilde R}_{41,41}(\l,\m)&=&
{\tilde R}_{61,61}(\l,\m)=
{\tilde R}_{71,71}(\l,\m)=
-\frac {(1+\l \m)(1+e^{2\eta}\l \m)
(1+e^{4\eta}\l \m)^2}
{e^{2\eta}(\l- \m)^2(e^{2\eta}\l- \m)
(e^{4\eta}\l -\m )
},\no\\
{\tilde R}_{42,42}(\l,\m)&=&
{\tilde R}_{43,43}(\l,\m)=
{\tilde R}_{62,62}(\l,\m)=
{\tilde R}_{65,65}(\l,\m)=
{\tilde R}_{73,73}(\l,\m)=
{\tilde R}_{75,75}(\l,\m)\no\\
&=&\frac {(1+\l \m)^2(1+e^{2\eta}\l \m)
(1+e^{4\eta}\l \m)^2}
{e^{\eta}(\l- \m)^2(e^{2\eta}\l- \m)
(\l-e^{2\eta}\m)
(e^{4\eta}\l -\m )
},\no\\
{\tilde R}_{44,44}(\l,\m)&=&
{\tilde R}_{66,66}(\l,\m)=
{\tilde R}_{77,77}(\l,\m)\no\\
&=&-\frac {(1+\l \m)^2(1+e^{2\eta}\l \m)
(1+e^{4\eta}\l \m)^2
(\l \m+e^{2\eta})}
{(\l- \m)^2(e^{2\eta}\l- \m)
(\l-e^{2\eta}\m)
(e^{4\eta}\l -\m)
(\l-e^{4\eta}\m)},\no\\
{\tilde R}_{45,45}(\l,\m)&=&
{\tilde R}_{63,63}(\l,\m)=
{\tilde R}_{72,72}(\l,\m)=
\frac {(1+\l \m)(1+e^{2\eta}\l \m)
(1+e^{4\eta}\l \m)}
{e^{2\eta}(\l- \m)^2
(e^{2\eta}\l-\m)
},\no\\
{\tilde R}_{45,63}(\l,\m)&=&
{\tilde R}_{63,72}(\l,\m)=
e^{\eta}{\tilde R}_{45,72}(\l,\m)\no\\
&=&\frac {(1+\l \m)(1+e^{2\eta}\l \m)
(1+e^{4\eta}\l \m)
((1+\l \m)
(1+e^{4\eta}\l \m )+(e^{4\eta}\l- \m)(\l- e^{2\eta} \m))}
{e^{3\eta}(\l- \m)^2(e^{2\eta}\l- \m)
(\l-e^{2\eta}\m)
(e^{4\eta}\l -\m)},\no\\
{\tilde R}_{45,81}(\l,\m)&=&
{\tilde R}_{81,72}(\l,\m)=
e^{-\eta}{\tilde R}_{63,81}(\l,\m)=
e^{-\eta}{\tilde R}_{81,63}(\l,\m)=
e^{-2\eta}{\tilde R}_{72,81}(\l,\m)=
e^{-2\eta}{\tilde R}_{81,45}(\l,\m)\no\\
&=&\frac {(1+\l \m)(1+e^{2\eta}\l \m)
(1+e^{4\eta}\l \m)
\sqrt{(1+\m ^2)(1+e^{4\eta}\l ^2)}}
{e^{2\eta}(\l- \m)^2
(e^{2\eta}\l -\m)
(e^{4\eta}\l -\m)},\no\\
{\tilde R}_{46,46}(\l,\m)&=&
{\tilde R}_{47,47}(\l,\m)=
{\tilde R}_{64,64}(\l,\m)=
{\tilde R}_{67,67}(\l,\m)=
{\tilde R}_{74,74}(\l,\m)=
{\tilde R}_{76,76}(\l,\m)\no\\
&=&-\frac {(1+\l \m)^2(1+e^{2\eta}\l \m)
(1+e^{4\eta}\l \m)}
{e^{\eta}(\l- \m)^2(e^{2\eta}\l- \m)
(\l-e^{2\eta}\m)}
,\no\\
{\tilde R}_{46,64}(\l,\m)&=&
{\tilde R}_{47,74}(\l,\m)=
{\tilde R}_{67,76}(\l,\m)\no\\
&=&-\frac {(1+\l \m)^2(1+e^{2\eta}\l \m)
(1+e^{4\eta}\l \m)
((1+e^{2\eta}\l ^2)
(1+e^{2\eta}\m ^2)+\l \m e^{2\eta}(1-e^{2\eta})(\l \m +e^{2\eta}))}
{(\l- \m)^2(e^{2\eta}\l- \m)
(\l-e^{2\eta}\m)
(e^{4\eta}\l -\m)
(\l-e^{4\eta}\m)},\no\\
{\tilde R}_{46,82}(\l,\m)&=&
{\tilde R}_{47,83}(\l,\m)=
{\tilde R}_{83,74}(\l,\m)=
{\tilde R}_{85,76}(\l,\m)\no\\
&=&e^{-\eta}{\tilde R}_{64,82}(\l,\m)
=e^{-\eta}{\tilde R}_{85,67}(\l,\m)=
e^{-2\eta}{\tilde R}_{67,85}(\l,\m)=
e^{-2\eta}{\tilde R}_{82,64}(\l,\m)\no\\
&=&e^{-3\eta}{\tilde R}_{74,83}(\l,\m)=
e^{-3\eta}{\tilde R}_{76,85}(\l,\m)=
e^{-3\eta}{\tilde R}_{82,46}(\l,\m)=
e^{-3\eta}{\tilde R}_{83,47}(\l,\m)\no\\
&=&-\frac {(1+\l \m)^2(1+e^{2\eta}\l \m)
(1+e^{4\eta}\l \m)
\sqrt{(1+\m ^2)(1+e^{2\eta}\l ^2)}}
{e^{\eta}(\l- \m)^2
(e^{2\eta}\l -\m)
(\l -e^{2\eta}\m)
(e^{4\eta}\l -\m)},\no\\
{\tilde R}_{48,48}(\l,\m)&=&
{\tilde R}_{68,68}(\l,\m)=
{\tilde R}_{78,78}(\l,\m)\no\\
&=&\frac {(1+\l \m)^2(1+e^{2\eta}\l \m)
(1+e^{4\eta}\l \m)
(\l \m+e^{2\eta})}
{(\l- \m)^2(e^{2\eta}\l- \m)
(\l-e^{2\eta}\m)
(\l-e^{4\eta}\m)},\no\\
{\tilde R}_{48,84}(\l,\m)&=&
{\tilde R}_{87,78}(\l,\m)=
e^{-2\eta}{\tilde R}_{68,86}(\l,\m)=
e^{-2\eta}{\tilde R}_{86,68}(\l,\m)
=e^{-4\eta}{\tilde R}_{78,87}(\l,\m)=
e^{-4\eta}{\tilde R}_{84,48}(\l,\m)\no\\
&=&\frac {(1+\l \m)^2(1+e^{2\eta}\l \m)
(1+e^{4\eta}\l \m)(\l \m +e^{2\eta})
\sqrt{(1+\l ^2)(1+\m ^2)}}
{(\l- \m)^2(e^{2\eta}\l- \m)
(\l-e^{2\eta}\m)
(e^{4\eta}\l -\m)
(\l-e^{4\eta}\m)},\no\\
{\tilde R}_{54,45}(\l,\m)&=&
{\tilde R}_{72,27}(\l,\m)=
\sqrt{(1+e^{2\eta}\l ^2)(1+e^{2\eta}\m ^2)}\no\\
& &\times\frac {(1+\l \m)(1+e^{2\eta}\l \m)
(1+e^{4\eta}\l \m)
((1+e^{4\eta}\l ^2)
(1+e^{4\eta}\m ^2)+(e^{4\eta}-1)(1+\l \m))}
{(\l- \m)^2(e^{2\eta}\l- \m)
(\l-e^{2\eta}\m)
(e^{4\eta}\l -\m)
(\l-e^{4\eta}\m)},\no\\
{\tilde R}_{63,45}(\l,\m)&=&
{\tilde R}_{72,63}(\l,\m)=
e^{-\eta}{\tilde R}_{72,45}(\l,\m)\no\\
&=&\frac {(1+\l \m)(1+e^{2\eta}\l \m)
(1+e^{4\eta}\l \m)
(e^{2\eta}(1+\l \m)
(1+e^{4\eta}\l \m )+(e^{4\eta}\l- \m)(\l- e^{2\eta} \m))}
{e^{\eta}(\l- \m)^2(e^{2\eta}\l- \m)
(\l-e^{2\eta}\m)
(e^{4\eta}\l -\m)},\no\\
{\tilde R}_{64,46}(\l,\m)&=&
{\tilde R}_{76,67}(\l,\m)=
e^{-2\eta}{\tilde R}_{74,47}(\l,\m)\no\\
&=&-\frac {(1+\l \m)^2(1+e^{2\eta}\l \m)
(1+e^{4\eta}\l \m)
(e^{2\eta}(1+e^{2\eta}\l ^2)
(1+e^{2\eta}\m ^2)+(e^{2\eta}-1)(\l \m +e^{2\eta}))}
{(\l- \m)^2(e^{2\eta}\l- \m)
(\l-e^{2\eta}\m)
(e^{4\eta}\l -\m)
(\l-e^{4\eta}\m)},\no\\
{\tilde R}_{81,81}(\l,\m)&=&
\frac {(1+\l \m)(1+e^{2\eta}\l \m)
(1+e^{4\eta}\l \m)}
{(\l- \m)(e^{2\eta}\l- \m)
(e^{4\eta}\l -\m )
},\no\\
{\tilde R}_{82,82}(\l,\m)&=&
{\tilde R}_{83,83}(\l,\m)=
{\tilde R}_{85,85}(\l,\m)=
-\frac {(1+\l \m)^2(1+e^{2\eta}\l \m)
(1+e^{4\eta}\l \m)}
{(\l- \m)^2(e^{2\eta}\l- \m)
(e^{4\eta}\l -\m )
},\no\\
{\tilde R}_{84,48}(\l,\m)&=&
{\tilde R}_{86,68}(\l,\m)=
{\tilde R}_{87,78}(\l,\m)=
\frac {(1+\l \m)^2(1+e^{2\eta}\l \m)
(1+e^{4\eta}\l \m)(\l \m +e^{2\eta})}
{(\l- \m)^2(e^{2\eta}\l- \m)
(\l-e^{2\eta}\m)
(e^{4\eta}\l -\m)},\no\\
{\tilde R}_{88,88}(\l,\m)&=&
-\frac {(1+\l \m)^2(1+e^{2\eta}\l \m)
(1+e^{4\eta}\l \m)(\l \m +e^{2\eta})(\l \m +e^{4\eta})}
{(\l- \m)^2(e^{2\eta}\l- \m)
(\l-e^{2\eta}\m)
(e^{4\eta}\l -\m)(\l -e^{4\eta}\m )}\no
\eea

Having found the matrix $\tilde{R}(\l,\m)$, we can proceed to solve the
RE for $K_+(u)$. The procedure 
is similar to that for $K_-(\l)$, as
sketched in the previous Appendix. The final result is given by (\ref{k+}).

\sect{On crossing-unitarity}

As a sideline, we show in this appendix that all R-matrices associated with 
finite-dimensional representations of the quantum affine algebra 
$U_q[{\cal G}^{(k)}]$ ($k=1,2$) for
generic $q$, where ${\cal G}$ is any bosonic Lie algebra, enjoy
the crossing unitarity property. We address, however, that the 
R-matrix appeared in the present paper does not belong to this class.

It is shown in \cite{Res90} that for any pair of finite dimensional
$U_q[{\cal G}^{(1)}]$-modules $V$ and $W$, 
the R-matrix $R^{VW}(z)$
satisfies the crossing-unitarity relations:
\bea
(((R^{VW}(z)^{-1})^{t_1})^{-1})^{t_1}&=&(\pi_V(q^{2\rho})\otimes 1_W)
    R^{VW}(zq^{2g})(\pi_V(q^{-2\rho})\otimes 1_W),\no\\
(((R^{VW}(z)^{-1})^{t_2})^{-1})^{t_2}&=&(1_V\otimes \pi_W(q^{-2\rho}))
    R^{VW}(zq^{-2g})(1_V\otimes\pi_W(q^{2\rho})).\label{cu-untwisted}
\eea
where $\rho$ is the half-sum of the positive roots of ${\cal G}$, and
$2g=(\psi,\psi+2\rho)$, with $\psi$ being the highest root of ${\cal
G}$.

We now extend the arguments of \cite{Res90} to R-matrices associated with
the twisted quantum affine bosonic
algebra $U_q[{\cal G}^{(2)}]$, and prove the similar crossing-unitarity
relations.

Let us first of all recall some facts about the twisted affine algebra
${\cal G}^{(2)}$. Let ${\cal G}_0$ be the fixed point subalgebra under
the diagram automorphism $\widehat{\tau}$ of ${\cal G}$ of order 2. We
can decompose ${\cal G}$ as ${\cal G}_0$ and a ${\cal
G}_0$-representation ${\cal G}_1$ of ${\cal G}$. 
Let $\t_0$ be the highest weight of the ${\cal G}_0$-representation 
${\cal G}_1$.
Following the usual convention, we denote the weight
of ${\cal G}^{(2)}$ by $\L\equiv (\lambda,\kappa,\tau)$, where $\lambda$ is a
weight of ${\cal G}_0$. With this 
notation the nondegenerate form $(~,~)$ induced on the weights
can be expressed as
\beq
(\L,\L')=(\lambda,\lambda')+\kappa\tau'+\kappa'\tau.
\eeq
The fundamental dominant weights of ${\cal G}^{(2)}$,
$\lambda_{(i)}~(0\leq i\leq r)$, can be shown to be 
\beq
\lambda_{(0)}=(\t_0,\t_0)\g,~~~\lambda_{(i)}=(\lambda_i,0,0)+
  2(\lambda_i,\t_0)\g,~~
  i=1,\cdots,r
\eeq
where $\lambda_i$ are fundamental dominant weights of ${\cal G}_0$
and $\g=(0,1,0)$. The distinguished dominant weight is
\beq
\widehat{\rho}=\sum_{i=0}^r\;\lambda_{(i)}=\rho+g_0\g
\eeq
where $g_0=(\t_0,\t_0+2\rho)$ and $\rho~(=\sum_{i=1}^r\;\lambda_i)$ 
is the half-sum of positive
roots of ${\cal G}_0$. 

We shall not give the defining relations for $U_q[{\cal G}^{(2)}]$,
but mention that the actions of coproduct and antipode on its 
generators $\{h_i,~e_i,~f_i,~d,~0\leq i\leq r\}$ are given by
\bea
\D(h_i)&=&h_i\otimes 1+1\otimes h_i,~~~~\D(d)=d\otimes 1+1\otimes
          d,\no\\
\D(e_i)&=&e_i\otimes q^{\frac{h_i}{2}}+q^{-\frac{h_i}{2}}\otimes
          e_i,~~~~
\D(f_i)=f_i\otimes q^{\frac{h_i}{2}}+q^{-\frac{h_i}{2}}\otimes
          f_i,\no\\
S(a)&=&-q^{\widehat{\rho}}\;a\;q^{-\widehat{\rho}},~~~~\forall 
   a=d,\;h_i,\;e_i,\;f_i.
\eea
Define an automorphism $D_z$ of $U_q[{\cal G}^{(2)}]$ by
\beq
D_z(e_i)=z^{\d_{i0}}e_i,~~~~~D_z(f_i)=z^{-\d_{i0}}f_i,~~~~
D_z(h_i)=h_i,~~~~D_z(d)=d.
\eeq
Then it can be shown that
\beq
S^2(a)=q^{2\rho}\,D_{q^{g_0}}\,(a)\,q^{-2\rho},~~~~~
       S^{-2}(a)=q^{-2\rho}\,D_{q^{-g_0}}\,(a)\,q^{2\rho}
\eeq
Following \cite{Res90}, we define the right dual module $V^*$ and
left dual module ${}^*V$ by
\beq
\pi_{V^*}(a)=\pi_V(S(a))^t,~~~~~~
\pi_{{}^*V}(a)=\pi_V(S^{-1}(a))^t,
\eeq
respectively.
By the same arguments as in \cite{Res90}, one can show that 
\beq
R^{V^*,W}(z)=(R^{VW}(z)^{-1})^{t_1},~~~~~
R^{V,{}^*W}(z)=(R^{VW}(z)^{-1})^{t_2}.
\eeq
It follows from the representations of $R^{V^{**},W}(z)$ and
$R^{V,{}^{**}W}(z)$ that for
any pair of finite dimensional $U_q[{\cal G}^{(2)}]$-modules $V$ and
$W$, the R-matrix satisfies the following crossing-unitarity relations
\bea
(((R^{VW}(z)^{-1})^{t_1})^{-1})^{t_1}&=&(\pi_V(q^{2\rho})\otimes 1_W)
    R^{VW}(zq^{g_0})(\pi_V(q^{-2\rho})\otimes 1_W),\no\\
(((R^{VW}(z)^{-1})^{t_2})^{-1})^{t_2}&=&(1_V\otimes \pi_W(q^{-2\rho}))
    R^{VW}(zq^{-g_0})(1_V\otimes\pi_W(q^{2\rho})).\label{cu-twisted}
\eea
Note also that
\beq
(\pi_V(q^{\pm 2\rho})\otimes\pi_W(q^{\pm 2\rho}))R^{VW}(z)
  =R^{VW}(z)(\pi_V(q^{\pm 2\rho})\otimes\pi_W(q^{\pm 2\rho})).\label{rho}
\eeq

Let us remark that if one uses the opposite coproduct and antipode of
$U_q[{\cal G}^{(2}]$,
\bea
\bar{\D}(h_i)&=&h_i\otimes 1+1\otimes h_i,~~~~\bar{\D}(d)=d\otimes 1+1\otimes
          d,\no\\
\bar{\D}(e_i)&=&e_i\otimes q^{-\frac{h_i}{2}}+q^{\frac{h_i}{2}}\otimes
          e_i,~~~~
\bar{\D}(f_i)=f_i\otimes q^{-\frac{h_i}{2}}+q^{\frac{h_i}{2}}\otimes
          f_i,\no\\
\bar{S}(a)&=&-q^{-\widehat{\rho}}\;a\;q^{\widehat{\rho}},~~~~\forall 
       a=d,\;h_i,\;e_i,\;f_i,
\eea
and denote the corresponding R-matrix by $\bar{R}(z)$, then the similar
arguments as above give rise to the following crossing-untarity
relations:
\bea
(((\bar{R}^{VW}(z)^{-1})^{t_1})^{-1})^{t_1}&=&(\pi_V(q^{-2\rho})\otimes 1_W)
    \bar{R}^{VW}(zq^{-g_0})(\pi_V(q^{2\rho})
    \otimes 1_W),\no\\
(((\bar{R}^{VW}(z)^{-1})^{t_2})^{-1})^{t_2}&=&(1_V\otimes \pi_W(q^{2\rho}))
    \bar{R}^{VW}(zq^{g_0})
    (1_V\otimes\pi_W(q^{-2\rho})).\label{cu-twisted-bar}
\eea



\begin{thebibliography}{99}
\bibitem{Skl88} E.K. Sklyanin, J. Phys. {\bf A:} Math.Gen. {\bf 21} (1988)
   2375.
\bibitem{Mez91} L. Mezincescu, R. Nepomechie, J. Phys. {\bf A:} Math.
   Gen. {\bf 24} (1991) L17; Int. J. Mod. Phys. {\bf A6} (1991) 5231.
\bibitem{deV93} H.J. de Vega, A. Gonz\'alez-Ruiz, J. Phys. {\bf A:}
   Math. Gen. {\bf 26} (1993) L519; Mod. Phys. Lett. {\bf A9} (1994)
   2207; \\
   J. Abad, M. Rios, Phys. Lett. {\bf B352} (1995) 92.
\bibitem{Yun95} C.M. Yung, M.T. Batchelor, Nucl. Phys. {\bf B435} (1995)
   430;\\ 
   M. Jimbo, R. Kedem, T. Kojima, H. Konno, T. Miwa, Nucl. Phys.
   {\bf 441} (1995) 437.
\bibitem{Zho96a} A. Gonz\'alez-Ruiz, Nucl. Phys. {\bf B424} (1994)
   553;\\
   H.-Q. Zhou, Phys. Rev. {\bf B54} (1996) 41.
\bibitem{Asa96} H. Asakawa, M. Suzuki, J. Phys. {\bf A:} Math. Gen. 
   {\bf 29} (1996) 225;\\
   M. Shiroishi, M. Wadati, J. Phys. Soc. Jpn. {\bf 66} 
   (1997) 1.
\bibitem{Zho96b} H.-Q. Zhou, Phys. Rev. {\bf B53} (1996) 5089.
\bibitem{Zha97}  Y.-Z. Zhang, H.-Q. Zhou, preprints cond-mat/9707263;
   cond-mat/9708003.
\bibitem{Bra97} A.J. Bracken, X.-Y. Ge, Y.-Z. Zhang, H.-Q. Zhou,
   preprint cond-mat/9710141.
\bibitem{Zho96R} H.-Q. Zhou, J. Phys. {\bf A:} Math. Gen. {\bf 29}
  (1996) 5509; Phys. Lett. {\bf A221} (1996) 104;\\
   H.-Q. Zhou, D.-M. Tong,  Phys. Lett. {\bf A232} (1997) 377. 
\bibitem{Bar91} R.Z. Bariev, J. Phys. {\bf A:} Math. Gen. {\bf 24}
   (1991) L919.
\bibitem{Res90} N.Yu. Reshetikhin, M.A. Semenov-Tian-Shansky, 
  Lett. Math. Phys. {\bf 19} (1990) 133;\\
  I.B. Frenkel, N.Yu. Reshetikhin, Commun. Math. Phys. {\bf 146} (1992)
  1.
\end{thebibliography}
\end{document}